\documentclass[twocolumn,pra,superscriptaddress]{revtex4-1}

\usepackage{epsfig}
\usepackage{graphicx}
\usepackage{amsmath}
\usepackage{amssymb}
\usepackage{epstopdf}
\usepackage{xcolor}
\usepackage{subfigure}
\usepackage{bm}
\usepackage{dsfont}
\usepackage{hyperref}
\usepackage{braket}
\usepackage{todonotes}
\usepackage{physics}
\usepackage{comment}
\newcommand{\mi}{\text{i}}

\begin{document}

\title{Photonic quantum memory using an intra-atomic frequency comb}

\author{G.~P.~Teja}
\affiliation{Indian Institute of Science Education and research, Mohali, Punjab 140306, India}

\author{Christoph~Simon}
\affiliation{Institute for Quantum Science and Technology and Department of Physics and Astronomy, University of Calgary, Canada T2N~1N4}

\author{Sandeep~K.~Goyal}
\email{skgoyal@iisermohali.ac.in}
\affiliation{Indian Institute of Science Education and research, Mohali, Punjab 140306, India}

\begin{abstract}
  Photonic quantum memory, such as an atomic frequency comb (AFC), is essential to make photonic quantum computation and long distance quantum communication scalable and feasible. In standard AFC the frequency of different atoms must be stable relative to each other which presents difficulties in realizing the quantum memory.  Here we propose  a quantum memory using an intra-atomic frequency comb which does not require frequency stabilization. We  show that the transitions between two degenerate energy levels of a single atom can be used to construct the frequency comb. The spacing between the teeth of the comb is controlled by applying an external magnetic field. Since the frequency comb is constructed from individual atoms,  these atoms can be used alone or in ensembles to realize the quantum memory.  Furthermore, the ensemble based quantum memory with intra-AFC is robust against Doppler broadening which makes it useful for high-temperature quantum memory. As an example, we numerically show the intra-AFC in cesium atoms and demonstrate a photon echo which is essential for quantum memory.

\end{abstract}

\maketitle
\section{Introduction}
Single photons are essential for long-distance quantum communication and linear optical quantum computation\cite{Beveratos2002,Reim2011,Zhang2008,Lukin2003,Sangouard2011,Northup2014,OBrien2009,Knill2001,Popescu2007,Kok2007}. The probabilistic nature inherent in conventional single-photon sources hamper the scalable implementation of such protocols~\cite{Lounis2000,Kurtsiefer2000,Santori2001,Senellart2017,Eisaman2011,Brunel1999}. Quantum memory, a device which can store and reemit single photons on demand, can help overcome this problem and allows efficient photonic quantum technologies~\cite{Heshami2016,Simon2010,Lvovsky2009,Brown2016}.

The basic idea behind a quantum memory is the light-matter interaction which allows the controlled reversible transfer of the quantum information between the photonic and the matter systems. Several protocols have been used to store single photons in atomic, condensed matter and superconducting systems~\cite{England2015,Grezes2015,Iakoupov2013,Specht2011,Veissier2013,Wolters2017,Reagor2016}. Electromagnetically induced transparency~\cite{Lukin2003,Longdell2005,Gorshkov2007}, controlled reversible inhomogeneous broadening~\cite{Nilsson2005,Kraus2006,Sangouard2007,Iakoupov2013}, and the atomic frequency comb (AFC)~\cite{Afzelius2009,Afzelius2010,Tittel2009,Jobez2016,Amari2010,Chaneliere2010,DeRiedmatten2008} are the most used quantum memories in the atomic ensembles. In all these quantum memories the incoming photon interacts with a carefully designed spectrum of the atomic ensemble. By controlling the shape and the characteristics of the spectrum one can store and retrieve photons from the ensemble.

\begin{figure*}
  \subfigure[]{
\includegraphics[height=5cm]{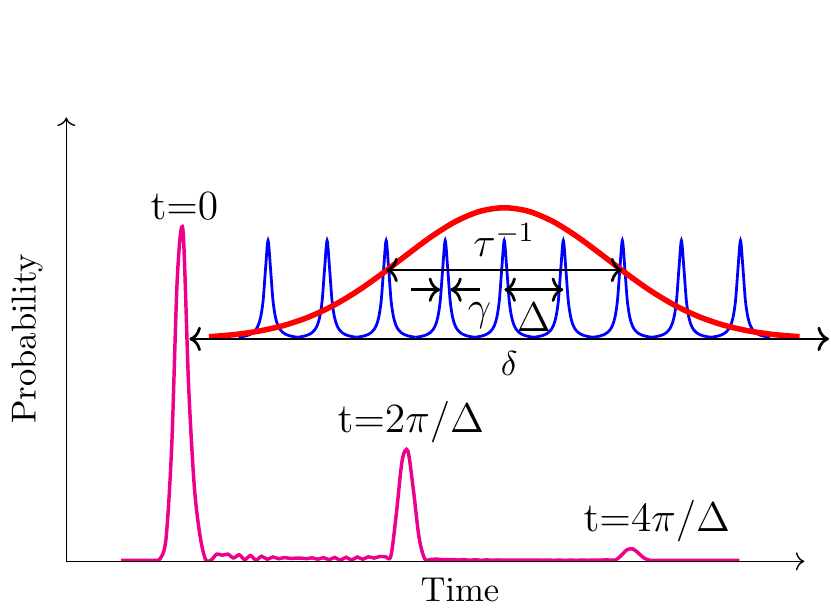}
\label{Fig:AFC}}
  \subfigure[]{
    \includegraphics[height= 3.5cm]{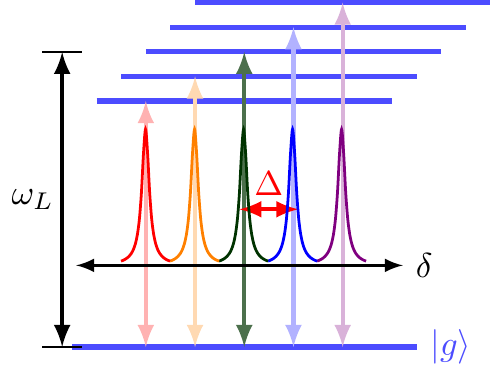}
    \label{fig:atom01}} 
\subfigure[]{
  \includegraphics[height= 3.5cm]{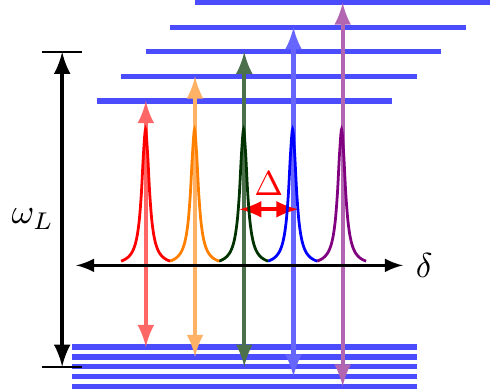}
\label{fig:atom02}}
  \caption{(Color online)\subref{Fig:AFC} Electric field output from an AFC. Here the first peak at time $t=0$ shows the probability (intensity) of the unabsorbed light. We observe photon echo at time interval of $2\pi/\Delta$. In the inset, we show the spectrum of the AFC.  The teeth in the AFC have width $\gamma$ and are separated by $\Delta$. Here, $\Gamma=N\Delta$ is the width of the comb  for $N$ number of teeth, $\delta$ is  the detuning, and $\tau$ is the temporal width of input laser pulse.  \subref{fig:atom01} An atom with a single ground state and multiple degenerate excited states. The degeneracy between the excited states is lifted by applying external magnetic field. $\omega_L$ is the mean transition frequency between the ground and the excited states and $\Delta \ll \omega_L$ is the spacing between different transitions. Each of the transition has a natural line-width $\gamma$. Hence, the transition spectrum of this atom forms an AFC. \subref{fig:atom02} Here we consider multiple ground and excited states with different allowed transitions. The mean transition frequency is $\omega_L$; however, the spacing between different transitions need not be the same. Therefore, the spectrum forms a non-uniform AFC.}
  \label{fig:atom}
\end{figure*}

Implementing AFC in an ensemble (in the usual way) relies on the frequencies of different atoms being stable relative to each other. This can be achieved in solids to some reasonable approximation at sufficiently low temperature. However, in gases Doppler broadening is a major limiting factor. 
To overcome this problem, in this paper we propose an intra-atomic frequency comb (I-AFC). We show that a frequency comb (FC) can be constructed by putting together multiple transitions between the hyperfine levels of a single atom. The spacing between the different transition frequencies is controlled by an external magnetic field. In order to show that such systems can be used for photonic quantum memories first we show the photon echo from an ensemble of such atoms. To achieve long storage time one can transfer the excitation from the excited state space to spin state space by applying a $\pi$ pulse.  As an example, we numerically show the FC in cesium (Cs) atoms. We show that an ensemble of Cs atoms with this FC results in a prominent photon echo which is robust against the Doppler broadening.

The I-AFC can be used alone or in ensemble  to realize efficient quantum memories. Since the FC is constructed from different transitions of individual atoms, the I-AFC in ensemble is also robust against environmental effects such as  Doppler broadening, which makes it useful for high-temperature quantum memory.   Another important advantage of the proposed scheme is that the  spacing between different teeth of the FC is controlled by the external magnetic field. Therefore, the finesse of the comb can be controlled by varying the magnetic field which can influence the efficiency and the photon-echo time of the quantum memory.

We have organized the article as follows: in Sec.~\ref{Sec:AFC} we present the details of atomic frequency combs based quantum memory. Section.~\ref{Sec:IAFC} contains our model of I-AFC. We present a toy model of I-AFC in Subsec.~\ref{SubSec:Toy}, generalization of this model for realistic atoms in Subsec.~\ref{SubSec:Multi} and calculations for I-AFC in Cs atoms in Subsec.~\ref{SubSec:Cs}. In Sec.~\ref{Sec:Doppler} we study the effect of Doppler broadening on the I-AFC. We conclude in Sec.~\ref{Sec:Conclusion}.

\section{Atomic frequency comb: an introduction}\label{Sec:AFC}
We start by introducing the conventional AFC. The AFCs  are typically consist of rare-earth ions doped in crystals that have optical transition between the ground state $\ket{g}$ and the excited state $\ket{e}$~\cite{Afzelius2009,Afzelius2010,Tittel2009,Jobez2016,Amari2010,Chaneliere2010,DeRiedmatten2008}. This transition has a narrow homogeneous bandwidth $\gamma$ and a large inhomogeneous bandwidth $\Gamma_{in}$ ($\Gamma_{in}\gg\gamma$). The transition $\ket{g}$-$\ket{e}$ is spectrally shaped such that the atomic density function consists of a series of equispaced narrow peaks (teeth), with spacing $\Delta$, resulting in a comb like structure in frequency modes, Fig.~\ref{Fig:AFC}.

A single photon  with  spectral width $\tau^{-1} = \bar\omega \gg \Delta$ is absorbed in the AFC system  at time $t=0$, which is stored as a collective excitation delocalized over all the teeth in the system. Formally, the state of the AFC, after absorbing a single photon, can be written as
\begin{align}
\Ket{\Psi}_{\text{AFC}} &= \sum_{j=1}^M \left(c_je^{\mi\delta_j t}\ket{\{e_j\}} \prod_{k\ne j} \ket{\{g_k\}}\right).\label{Eq:AFC-Dicke}
\end{align} 
Here $\ket{\{g_j\}} \equiv \ket{g_1\,g_2\cdots\,g_{N_j}}_j$  and $\ket{\{e_j\}} \equiv \sum_n\alpha_n\ket{g_1\cdots e_n\cdots g_{N_j}}_j$ represent the ground and collective single-excitation state of all the atoms with detuning $\delta_j$, respectively, and the $c_j$'s represent the absorption coefficient of each tooth in the comb. The coefficients $\alpha_n$ characterize the absorption by individual atoms.

The photon emission probability $P(t)$ from this setup is proportional to
\begin{align}
  P(t) \propto |\bra{G}S_-\Ket{\Psi}_{\text{AFC}}|^2,\label{Eq:P}
\end{align}
where $S_- = \sum_j\Ket{\{g_j\}}\Bra{\{e_j\}}$ is the step down operator and $\ket{G} = \prod_k\ket{\{g_k\}}$ is the collective ground state of the ensemble. Here the sum is over all the teeth. It can be seen from the state~\eqref{Eq:AFC-Dicke} and the expression for the photon emission probability $P(t)$~\eqref{Eq:P} that the probability is maximum at times $t$ which are integer multiples of $2\pi/\Delta$ [Fig.~\ref{Fig:AFC}]. The unabsorbed light comes out at $t=0$, whereas at time $2n\pi/\Delta$ we get the $n$th photon echo.

The efficiency $\eta$ of the quantum memory in this protocol  is defined as the  ratio of the total amount of light coming out in the first echo (at time $t=2\pi/\Delta$) and the total input intensity, i.e., 
\begin{align}
\eta=\frac{\int_{\pi/\Delta}^{3\pi/\Delta} \abs{E_{\text{out}}(t)}^2 \,dt}{\int\abs{E_{\text{in}}(t)}^2 \,dt}.
\end{align}
Theoretically, the maximum efficiency that can be achieved in this manner is $54\%$~\cite{Afzelius2009,Afzelius2010}. To store the excitation for a long time the excitation is transferred to a long-lived spin state by applying a $\pi$ pulse. This excitation can be retrieved at a later time by applying another $\pi$ pulse which transfers the excitation from the spin state to the excited state from where we can observe a photon echo at time $2\pi/\Delta$. The application of two $\pi$ pulses impart an overall negative phase in the system which causes the photon to emit in the backward direction. In this scenario the efficiency of the photon echo can approach $100\%$~\cite{Afzelius2009,Afzelius2010}.


As mentioned earlier, the usual ways of implementing AFC require the frequencies of different atoms in the ensemble to be stable relative to each other which is often difficult to achieve in a gaseous ensemble. 
In the following we present a model where the FC is constructed from individual atoms which do not have the above said problems. 

\section{Intra-atomic frequency comb}\label{Sec:IAFC}
In this section we introduce the concept of I-AFC. We first present a toy model in Sec.~\ref{SubSec:Toy} where the atom has only a single ground state and multiple excited states. By allowing the transitions only between the excite and the ground states we show that each such atom is capable of producing a photon echo, exactly like the one in AFC. We generalize this result for more realistic systems in Sec.~\ref{SubSec:Multi}. In Sec.~\ref{SubSec:Cs} we consider the cesium atom as an example and numerically show a photon echo.

\subsection{I-AFC: A toy model}\label{SubSec:Toy}
Consider an atom with a single ground state and $N$ the number of degenerate excited states~[Fig.~\ref{fig:atom01}]. We lift the degeneracy in the excited state space by applying an external magnetic field (Zeeman effect). The excited states are chosen such that 
the transition is possible only between the ground state and the excited states. The mean transition frequency is $\omega_L$, and the spacing between different transitions is $\Delta$; hence, the transition spectrum for this atom will resemble an AFC. The question is whether a photon echo can be observed in this atom in order to be eligible for quantum memory.

The Hamiltonian (in the interaction picture) for this atom in the presence of an electromagnetic field reads~(see Appendix~\ref{ape} for details)
\begin{equation}
H =\sum_{n=1}^{N}\hbar \delta_{n} \dyad{n}{n}-\dfrac{\mathcal E(z,t)\,d}{2}\sum_{n=1}^{N}(\dyad{n}{0}+\dyad{0}{n}), \label{h}
\end{equation} 
where $\delta_{n}$ is the detuning between the mean frequency $\omega_L$ of light and the frequency of the $n$th excited state $\Ket{n}$, $\ket{0}$ is the ground state, and $d$ is the magnitude of the transition dipole operator which we have taken to be same for all the transitions. $\mathcal E(z,t)$ represents the amplitude of the electric field.

If the state of the atom at time $t$ is given as the superposition of all the states, i.e., 
$\ket{\psi(t)}= \sum _{n=0}^{N}c_{n}(t)\ket{n}$,
 then the dynamical equation in terms of the coefficients $c_n(t)$ reads
\begin{align}
\dot{c}_{n}=&  \dfrac{\mi d}{2 \hbar} \mathcal E(z,t)  c_{0}-\mi\delta_{n}c_{n},\quad \dot{c}_{0}= \dfrac{\mi d}{2 \hbar} \mathcal E(z,t) \sum_{n=1}^{N} c_{n}(t).\label{s2}
\end{align}
The interaction between the photon and the atom is weak. Hence, the probability of finding the atom in the ground state is almost unity, i.e, $|c_0|^2 \simeq 1$. Therefore, 
\begin{align}
c_{n}(t)=&\dfrac{\mi d}{2\hbar}e^{-\mi \delta_{n} t}\int_{-\infty}^{t}e^{\mi \delta_{n} t'} \mathcal E(z,t') \,dt'. \label{s1}
\end{align}
For a  Gaussian input, i.e.,  electric field of the form $\mathcal{E}(z,t)=\mathcal{E}_{0}\exp(-t^2/2\tau^2)$ with the temporal  width  $\tau \ll t$ Eq.~(\ref{s1}) yields
\begin{align}
c_{n}(t)=&\dfrac{ \mi \Omega}{2} \dfrac{\sqrt{\pi}}{\bar{\omega}} \exp[-(\delta_{n}\tau)^{2}/2-\mi \delta_{n} t], \nonumber
\end{align}
where $\Omega=\dfrac{ d \mathcal{E}_{0}}{\hbar}$.

Since we have assumed equispaced excited states, we can  write  $\delta_n = \delta_0 + n \Delta$ for some constant $\delta_0$. The photon emission probability $P(t)$ (for forward propagation) for this atomic system can be redefined as
\begin{align}
  P(t) =|\bra{0} s_-\ket{\psi(t)}|^2 =  \left|\dfrac{ \mi \Omega}{2} \dfrac{\sqrt{\pi}}{\bar{\omega}} \sum_{n=1}^{N} e^{-\mi nt \Delta }\right|^2,\label{Eq:photon-prob}
\end{align}
 where $s_- = \sum_{n} \Ket{0}\Bra{n}$  and the sum is over all the states in the excited level.
Here we have assumed that $\bar{\omega} =\tau^{-1}\gg \Delta$ which implies that $ \exp[-(n\Delta)^2/2\bar{\omega}^{2}] \to 1$. Clearly, Eq.~\eqref{Eq:photon-prob} results in a photon echo at $t = 2\pi/\Delta$ when the probability of the photon emission is maximum. More detailed calculations can be found in Appendix~\ref{ape}.

\begin{figure*}[t!]
  \subfigure[]{
    \includegraphics[width= 6cm]{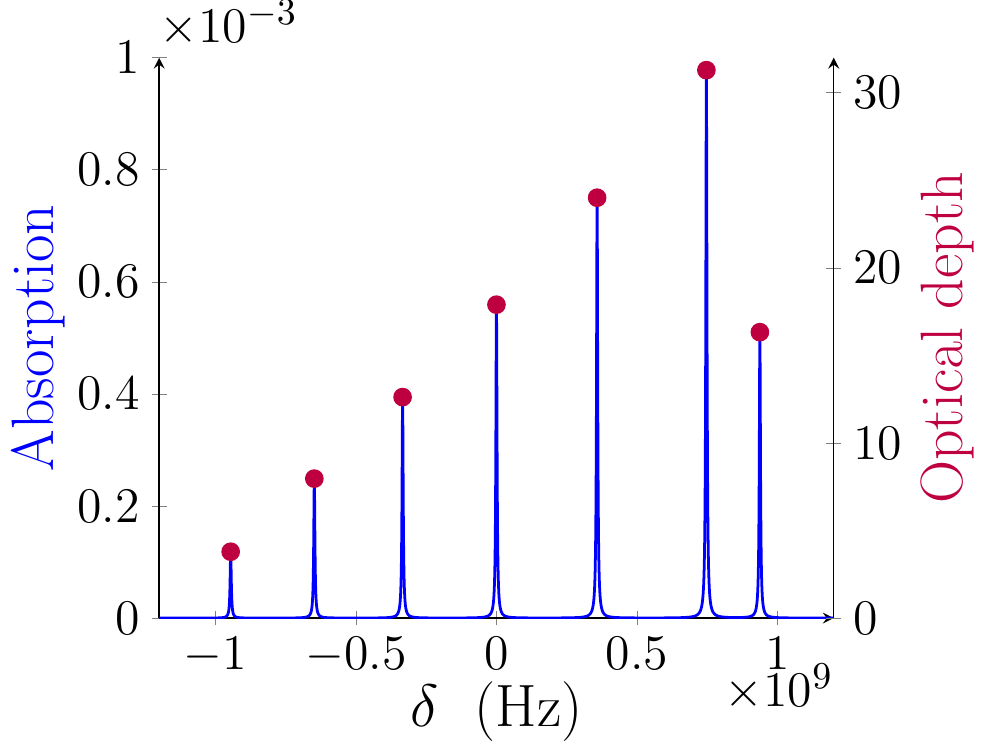}
    \label{fig:m+1}} 
\subfigure[]{
  \includegraphics[width= 6cm]{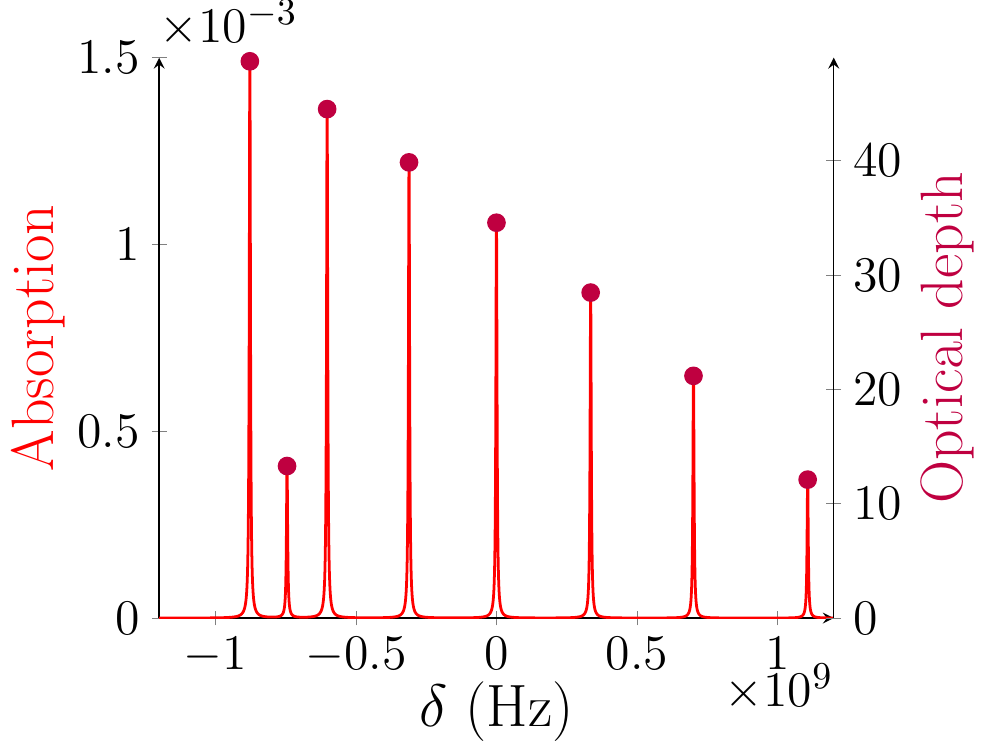}
\label{fig:m-1}}\\
\subfigure[]{
  \includegraphics[width= 6cm]{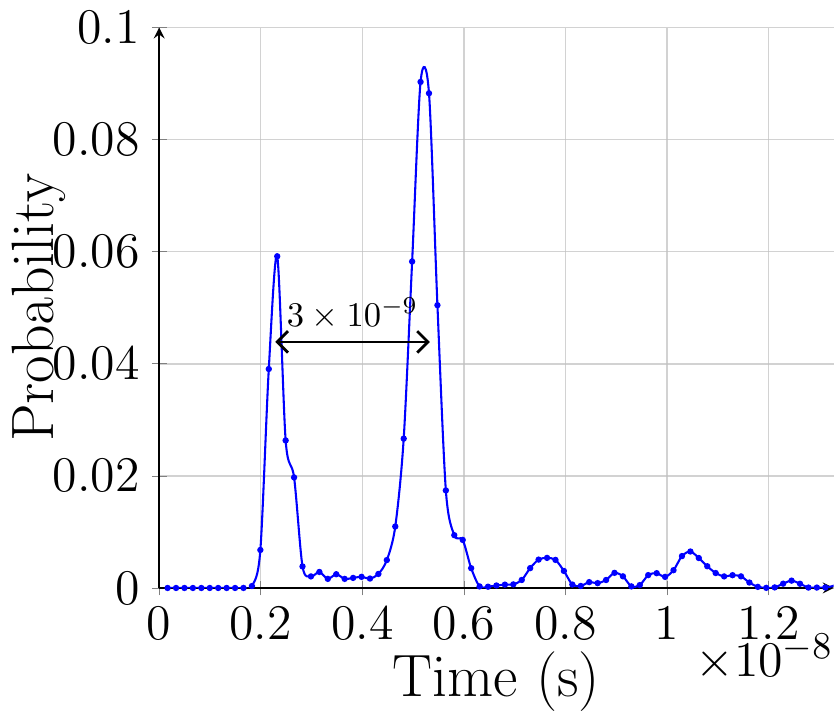}
\label{fig:Output-4}}
\subfigure[]
{\includegraphics[width= 6cm]{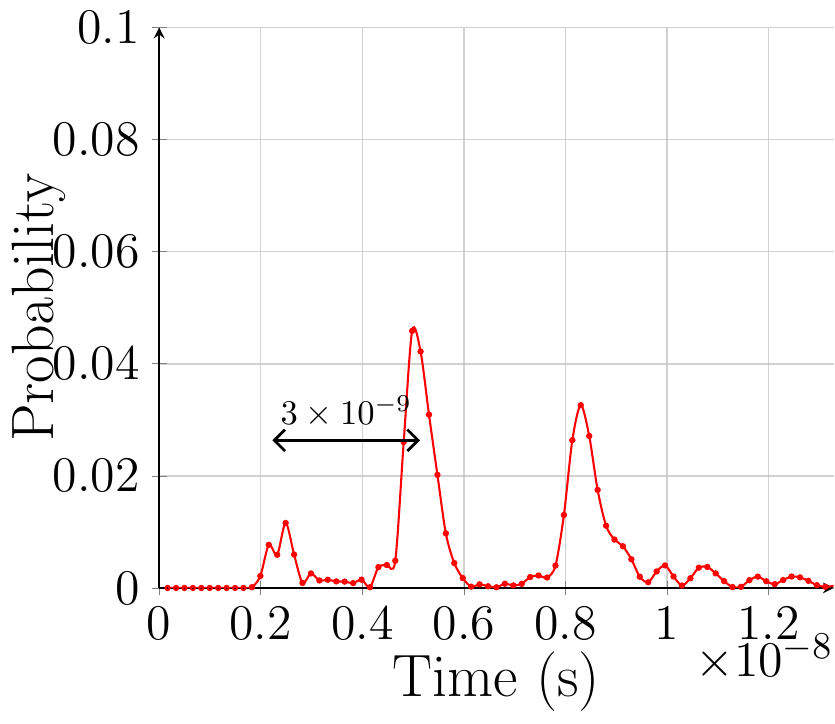}
\label{fig:Output-5}}
\caption[]{(Color online) Nonuniform AFC  and the photon echo in a Cs atom. \subref{fig:m+1} and \subref{fig:m-1}: Numerically calculated absorption and optical depths for various  transitions for  $\Delta m = + 1$ and $\Delta m = -1$. Here the $x$ axis is the detuning with respect to the incoming light and the $y$ axis is the value of the absorption for various transitions. The solid circles represent the optical depth corresponding to each transition for a density of $10^{18}m^{-3}$, length of the vapour cloud cell $L = 5cm$ and the natural line-width $\gamma = 5$MHz.
  \subref{fig:Output-4}, \subref{fig:Output-5}: The photon echo observed in Cs atoms for $\Delta m = \pm 1$ transitions. }
\label{fig:Cesium}
 \end{figure*}


 This calculation confirms that an ideal I-AFC is capable of demonstrating photon echo, which can  ultimately be used as a quantum memory. However, in a real atom the situation can be slightly different. Generally, atoms have a number of degenerate states in both the ground and the excited levels~\cite{Sansonetti2006,Sansonetti2009} which can make calculations a little difficult. In the next section we consider a realistic model for the atom with multiple ground and excited states and prove that they can produce a photon echo.

\subsection{Multilevel atom: Multiple ground and excited states}\label{SubSec:Multi}

Here we solve for the dynamics of an atom containing $N_g$ number of ground states $N_e$ number of excited states and  $N_s$ number of spin states. These states are chosen such that the dipole transition is allowed between the ground and excited states and excited and spin states. Whereas the transition between the ground and the spin states is forbidden. To realize a quantum memory, first the atom is prepared in the uniform superposition of ground states. Upon absorption of a single photon the atom will acquire a state which is a superposition of various excited states. To make this storage of a single photon, we apply a $\pi$ pulse which transfers the excitation from the excited level to the spin level. In order to show that an I-AFC is capable of storing a single photon first we show the calculations for the photon echo. These calculations will be followed by the calculations showing a near perfect transfer of the excitation from excited states to spin states and vice versa.

{\em Photon echo:} Since the spin level does not participate in the photon echo, we consider only the ground and the excited levels which are interacting with  oscillating electric field $E$. The Hamiltonian for such an system reads
\begin{widetext}
\begin{align}
H =\sum_{n=1}^{N_e} \hbar\omega_{n}^{e}\dyad{e_n}{e_n} +\sum_{m=1}^{N_g}\hbar\omega_{m}^{g} \dyad{g_m}{g_m} -\dfrac{\mathcal E(z,t)}{2}\qty(\sum_{n,m}d_{nm}\dyad{e_n}{g_m} e^{-\mi \omega_L t}+H.C), \label{mgh}
\end{align}
\end{widetext}
where $\ket{e_n}$ ($\ket{g_m}$) represents the $n$ ($m$th) excited (ground) state corresponding to energy $\hbar \omega^e_{n}$ ($\hbar \omega^g_{m}$), $\ket{g_{_1}}$ being the ground state with zero energy,
and $d_{nm}$ is the transition dipole moment between $\ket{e_{_n}}$ and $\ket{g_{_m}}$. Here we are assuming that the transition among the ground states and among the excited states is forbidden. Therefore, $d_{nn^\prime}=d_{mm^\prime}=0$.

The Hamiltonian for this atom in the interaction picture can be calculated as $H_I = e^{\mi H_{0} t} H e^{-\mi H_{0} t} -H_0$ using $H_{0}= \sum_{n=1}^{N_e}\hbar \omega_{L} \dyad{e_n}{e_n}$. The Hamiltonian $H_I$ reads\begin{widetext}      
\begin{align}
  H_I &=\sum_{n=1}^{N_e} \hbar (\omega_n^e-\omega_L)\dyad{e_n}{e_n} +\sum_{n=1}^{N_g}\hbar\omega_{n}^{g} \dyad{g_n}{g_n}-\dfrac{\mathcal E(z,t)}{2}\qty(\sum_{n,m}d_{nm}\dyad{e_n}{g_m} +H.C), 
\end{align}
\end{widetext}
Similar to the previous case of single ground state and multiple excited states, we can assume that the population in the ground state is much smaller than the excited states, i.e.,  $\sum_m\rho_{mm}\simeq 1$  and $\rho_{nn}\simeq\rho_{nn^\prime}\simeq \rho_{mm^\prime}\simeq 0$ but $\rho_{nm}\neq 0$. This yields the dynamical equation for the coherence
\begin{align}
\pdv{{\rho_{nm}(z,t)}}{t}+\qty(\mi \delta_{nm}+\dfrac{\gamma}{2})\rho_{nm}(z,t)=\dfrac{\mi d_{nm}}{2\hbar}\rho_{mm} \,\mathcal E(z,t),\label{dme}
\end{align}
where $\delta_{nm}=[(\omega^e_{_n}-\omega^g_{_m})-\omega_L]$ is the detuning between the transition $\ket{e_n}\leftrightarrow\ket{g_m}$ and the mean frequency of the laser light.\par

Using the definition for the atomic polarization
\begin{align}
\mathcal{P}(z,t)=2\mathcal{N} \sum_{n,m} d^{*}_{nm}\rho_{nm},
\end{align}
we can arrive at the equations
{\begin{align}
\tilde{\rho}_{nm}(z,\omega)&=\dfrac{\mi d_{nm} \, \rho_{mm} } {2\hbar} \dfrac{\tilde{\mathcal{E}}(z,\omega)}{\mi(\delta_{nm}+\omega)+\gamma/2},\\
\mathcal{P}(z,\omega)&=2 \mathcal{N} \sum_{n,m} \dfrac{\mi \abs{d_{nm}}^{2}\rho_{mm}}{2\hbar}\dfrac{\tilde{\mathcal E}(z,\omega)}{\gamma/2+\mi(\delta_{nm}+\omega)} ,\label{fp}
\end{align}}
and propagation equation for $\mathcal{\tilde{E}}(z,\omega)$,
\begin{align}
\tilde{\mathcal{E}}(z,\omega)=\tilde{\mathcal{E}}(0,\omega)e^{-\mathcal{D}z}, \label{leq}
\end{align}
where $\mathcal{D}$ and $g$ are defined as
\begin{align}
\mathcal{D} = \sum_{n,m} \dfrac{g_{nm}}{\gamma/2+\mi(\delta_{nm}+\omega)}+\dfrac{\mi \omega}{c}, \qquad g_{nm}=\mathcal{N} \dfrac{\abs{d_{nm}}^{2}\rho_{mm} \,\omega_{L}}{2\hbar\epsilon c}.\nonumber 
\end{align}
Clearly $\mathcal{D}$ in this case is identical to the one we obtained for the case of single ground state. Mathematically both the cases are the same hence we get an echo in these generalized atoms.

{\em Action of the $\pi$ pulse:}
For simplicity, we assume that $N_e = N_s$, i.e., the number of the excited states is same as the number of spin states. Furthermore, we assume that the comb structure of the excited states is identical to that of the spin-state frequency comb.  upon the application of a strong driving field $E_d$ with mean frequency $\omega_d$, the Hamiltonian of the system in the interaction picture is given as
\begin{align}
H=-\dfrac{d\cdot E_d}{2} \sum_{n} \op{e_n}{s_n} + \op{s_n}{e_n} ,\label{sp1}
\end{align} 
where we have chosen $H_0 = \hbar \sum_{n =1}^{N_e} \omega_{n}^{e} \dyad{e_n} + \hbar \sum_{n}^{N_s} \omega_{n}^{s} \dyad{s_n}$ and since we assumed spin state and excited have the same comb we have used $\omega_{n}^{e} - \omega_{d}= \omega_{n}^{s}$.    The spin states are represented by $\Ket{s_n}$ with energy $\hbar \omega^s_n$ and dipole moment $d$ is taken same for all the transition between the excited and the spin states. The evolution of the state of this atom is captured by the operator $U(t_0,t) =\exp[-\text{i} H (t-t_0)]$. For the $\pi$ pulse we have $\int_0^t \Omega \, dt^\prime=\pi$ where $\Omega = d\cdot E_d$. This results in $U = -\text{i}\sum_{n}\left( \op{e_n}{s_n} + \op{s_n}{e_n}\right) $ which results in a perfect transfer of the excited state to the spin state. Next we consider the example of Cesium atom and show numerically that a photon echo can be observed in this system.
\subsection{I-AFC in Cesium atoms}\label{SubSec:Cs}
 For example, consider the Cs atom. Here a $16$-fold degenerate $5p^66s$ energy level is chosen as the ground level and a $32$-fold degenerate  $5p^68p$ level constitutes the excited state. The mean transition frequency between the two levels is $773.21$THz~\cite{Liu2000,Safronova1999}. The natural line-width of all the allowed transitions is taken to be equal ($5$MHz)~\cite{Liu2000}. The degeneracy in the ground-levels and the excited-levels is lifted by applying an external magnetic field of $0.1$T.

 In Figs.~\ref{fig:m+1} and~\ref{fig:m-1} we calculate the  most prominent absorption amplitudes and the optical depths between the chosen energy levels for $\Delta m = \pm 1$ transitions for an ensemble of Cs atoms in the vapor form with density $10^{18} m^{-3}$ and  the vapor cloud cell of length $5$ cm (see Appendix~\ref{App:Transitions}). The spectrum in these figures resemble the AFC but with nonuniform spacing between the teeth and with unequal heights.

 \begin{figure}
   \includegraphics[width=8cm]{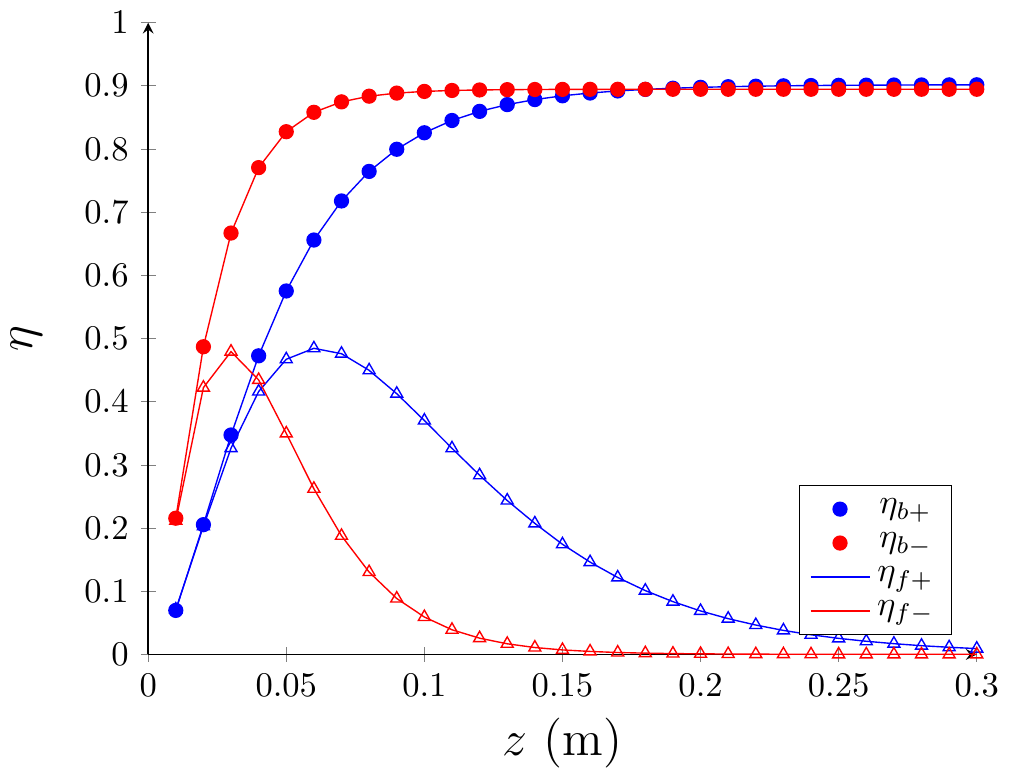}
   \caption{(Color online) Here we plot the photon echo efficiency for backward  $\eta_{b\pm}$ and forward  $\eta_{f\pm}$ emission for $\Delta m = \pm1$ transitoins. The maximum efficiency for teh forward emission is $\eta_f \sim 48\%$ and for the backward emission is $\eta_b \sim 90\%$.
   }\label{Fig:Efficiency}
 \end{figure}
 
In order to observe a photon echo, initially all Cs atoms are prepared in a uniform superposition of all the ground states using radio frequency optical pumping~\cite{Bloom1960}. Upon absorption of a single photon the atom will acquire the state which is a superposition of different excited states.
In Figs.~\ref{fig:Output-4} and~\ref{fig:Output-5} we show numerically the photon emission probability $P(t)$ for the forward propagation. The first peak at $t=2$ ns in both the plots  is the emission probability of the unabsorbed photon. The second peak is the photon echo which occurs $3$ ns after the input photon.

 In order to achieve a long time quantum memory we need to transfer the delocalized excitation from the excited level to a long-lived spin level. This can be done by applying a $\pi$ pulse between the excited level and the spin level. If the FC in the excited level is identical to the one in the spin level then theoretically $100\%$ transfer can be achieved. The population transfer can also be achieved by chirping of femtosecond pulses~\cite{Zhdanovich2008}. However, for the non identical FCs numerical optimization is required to achieve the maximum transfer. For the case of Cs the degenerate energy level  $5p^6 5d$ can serve as the long-lived spin state where the photon can be stored (in principle)indefinitely. To retrieve the photon we apply another $\pi$ pulse and these two pulses together cause the backward propagation of the photon.

In Fig.~\ref{Fig:Efficiency} we plot the photon-echo efficiency $\eta$ for backward and forward emissions as a function of the cell length $z$. The forward efficiency $\eta_{f\pm}$ satisfy the relation
\begin{align}
  \eta_{f\pm} \propto (\alpha_\pm z)^2\exp(-\alpha_\pm z),
\end{align}
where $\alpha_\pm$ are the optical depths for $\delta m = \pm1$ transitions. We found numerically that $\alpha_+= 32.09$ and $\alpha_- = 66.77$~(see Appendix~\ref{App:Efficiency}). The efficiency for the backward emission $\eta_{b\pm}$ follow the relation
\begin{align}
  \eta_{b\pm} \propto [1-\exp{-\alpha_\pm z}]^2.
\end{align}
The maximum efficiency $\eta_f$ we achieve is close to $48\%$ and for $\eta_b$ close to $90\%$ which is numerically optimized over the spectral width of the incoming photon. Note that the efficiencies are different for different polarizations, i.e., for different $\Delta m$.  The anisotropy in the two types of transition is due to the external magnetic field which causes non-uniform superposition of the hyperfine levels for $\Delta m = \pm 1$~\cite{Steck}.

 Note that both the ground states and the excited states have FC-like structure which might create complications; however, we find that the FC in the ground state does not affect the photon echo.

 So far we have shown that an I-AFC  can result in photon echo. Our results are supported by numerical calculations for Cs atoms. Next we show that the I-AFC is robust against environmental effect, particularly, Doppler broadening.

\uppercase{\section{Effect of Doppler broadening on I-AFC}} \label{Sec:Doppler}
 In an ensemble of Cs atoms in a thermal equilibrium at temperature $T$ the state of each of the atoms is $\ket{v}\otimes \ket{\psi_v}$ where $\ket{v}$ is the kinetic state of the atom and $\ket{\psi_v}$ is its electronic state labeled by the velocity $v$. The total state of the atomic ensemble can be written as
\begin{align}
  \rho = \int \text{d}v\,p(v) \ket{v}\bra{v}\otimes \ket{\psi_v}\bra{\psi_v},
\end{align}
where $p(v) = \exp(-mv^2/2k_BT)$ is the thermal distribution with $m$ being the mass of the atom and $k_B$ being the Boltzmann constant. The motion of the atom makes it perceive the frequency of the incoming photon shifted from $\omega_L$ to $\omega_L+\bm{v\cdot k}$ where $\bm{k}$ is the wave vector of the photon. Or it can be perceived as a shift of $-\bm{v}\cdot\bm{k}$ frequency in the FC. Since the time of the photon echo depends only on the separation $\Delta$ between the teeth of the FC,  it remains invariant under any such shift of the comb. Therefore, the photon echo from Cs atoms with different velocities would occur exactly at the same time $2\pi/\Delta$.

It can be understood as follows: the shift in the FC would result in an additional phase $\exp(\mi \bm{v}\cdot\bm{k} z/c)$ in the state, i.e.,  $\ket{\psi_v(t)} \to \text{e}^{\text{i}\bm{v\cdot k}z/c}\ket{\psi_v(t)}$~(see Appendix~\ref{App:Doppler}). However, since the thermal distribution is incoherent, this additional phase does not cause any interference. The photon-emission probability $P(t)$ for this system reads
\begin{align}
  P(t) = \int_v \text{d} v\, p(v) |\bra{G}s_-\ket{\psi_v(t)}|^2,
\end{align}
which is independent of the phase $\exp(\mi \bm{v}\cdot\bm{k} z/c)$. Here $\ket{G}$ is the collective ground state.
Therefore, the photon echo is the thermal average of the echos from all the atoms having different velocities and 
appear at exactly the same time. Hence, Doppler broadening has no effect on the I-AFC.

\begin{figure}
\includegraphics[width=8cm]{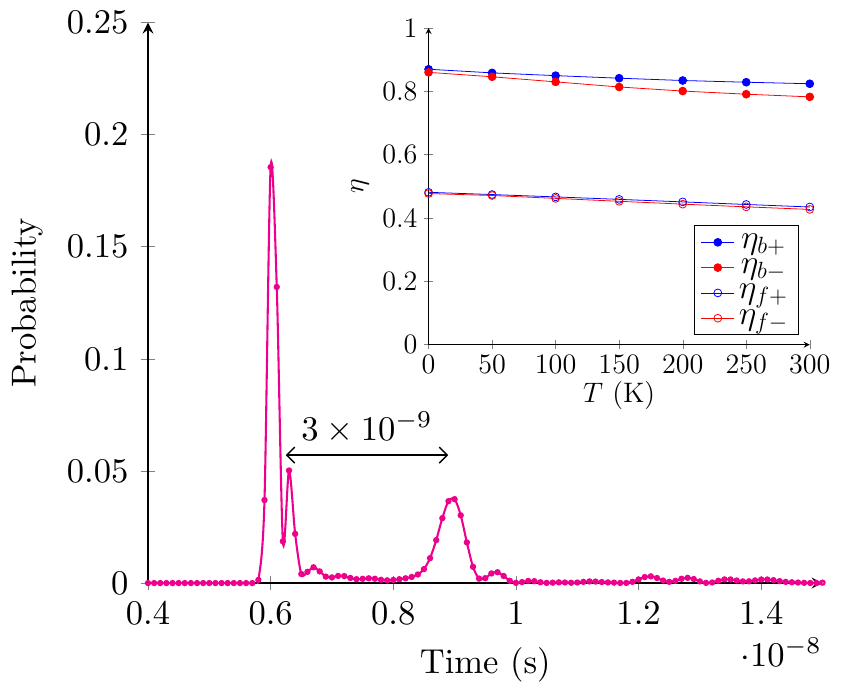}
\caption{(Color online) In this figure we show the photon echo which occurs $3$ns after the input pulse in the presence of Doppler broadening at $100$K temperature. In the inset, we show the photon echo efficiencies for the forward and for the backward propagation as a function of temperature. }\label{Fig:Doppler}
\end{figure}

In Fig.~\ref{Fig:Doppler} we plot the output from the Cs vapors at $100$K. In the inset of this figure we have the numerically optimized efficiencies for the backward and the forward propagation for $\Delta m = \pm 1$ as a function of temperature. The backward efficiency starts with a value close to $90\%$ at zero temperature and ends above  $80\%$ for $300$ K for both the transitions. Similarly, the forward efficiency is unaffected by temperature. This shows that the proposed scheme for quantum memory is robust against thermal effects.


\section{Conclusion}\label{Sec:Conclusion}
To conclude, we have proposed an I-AFC which is constructed using different transitions of a single atom.
This AFC is important for trapped ions and  on-chip quantum computation, single-atom quantum memory and microwave-to-optical transducers.  Since the I-AFC is constructed from individual atoms, this FC (in atomic ensembles) is immune against Doppler broadening; and hence can be used for high-temperature photonic quantum  memory. The spacing between the neighboring teeth in the comb which characterize the time of the photon echo is controlled by an external magnetic field. Therefore, the time of the photon echo can be controlled by tuning the magnetic field. Since the I-AFC can absorb both left-handed and right-handed polarized light ($\Delta m = \pm 1$), the proposed quantum memory is the most suitable for the polarization qubits. The anisotropy in the absorption of the two polarizations can be minimized by reducing the magnetic field. Any remaining imbalance in the absorption can be corrected by external polarization dependent loss. We considered the example of a Cs atom and showed that I-AFC can be realized in Cs atoms.

\begin{acknowledgments}
We thank K.~Heshami and M.~Singh for their useful comments. C.~S.~acknowledges funding from the Natural Sciences and Engineering Research Council (NSERC). S.~K.~G. acknowledges the financial support from SERB-DST (File No. ECR/2017/002404).
  
\end{acknowledgments}

\appendix

\uppercase {\section{Multilevel atom: single ground state and multiple excited states}}\label{ape}
Here  we solve for the dynamics of an atom containing a single ground state and $N$ number of excited states interacting with oscillating electric field $E$. The Hamiltonian for such an atom can be written as~\cite{bransden}
\begin{equation}
H =\sum_{n=1}^{N}\hbar \omega_{n0} \dyad{n}{n}-\dfrac{\mathcal E(z,t)}{2}\sum_{n=1}^{N}\qty( d_{n0} \dyad{n}{0} e^{-\mi \omega_{L} t}+ \text{h.c.}) ,\label{ham}
\end{equation}
where  $\Ket{n}$ represents the $n$-th excited state corresponding to energy  $\hbar\omega_{n0}$, $\Ket{0}$ being the ground state with zero energy,  and  $d_{n0}$ is the transition dipole moment between  $\Ket{0}$ and $\Ket{n}$. Here we are assuming that the transition between the different excited states is forbidden. Therefore,  $d_{nn'} =  0$ for $n,n'\ne 0$. We can write the oscillating electric field as
\begin{align}
E(z,t)&=\mathcal{E}(z,t)\cos(\omega_{L}t-kz), \label{Eq:efield}
\end{align}
where $\mathcal E(z,t)$ gives the temporal shape of the electric field and $\omega_L$ is the mean frequency of light.

The Hamiltonian $H$ in the interaction picture can be calculate as $H_I = e^{\mi H_{0} t} H e^{-\mi H_{0} t} -H_0$ using $H_{0}= \sum_{n=1}^{N}\hbar \omega_{L} \dyad{n}{n}$. The Hamiltonian $H_I$ reads 
\begin{align}
  H_I &=\sum_{n=1}^{N}\hbar \delta_{n} \dyad{n}{n}-\dfrac{\mathcal E(z,t)}{2}\sum_{n=1}^{N}\qty( d_{n0} \dyad{n}{0}+d^*_{n0} \dyad{o}{n}),
\end{align}
where  
$\delta_{n}=\omega_{n0} - \omega_{L}$ is the detuning between the $n$-th energy level and the mean frequency of the laser light.

The state of the atomic system is represented by the density operator $\rho$ where $\rho_{nn}$ is the population of the $n$th energy level and $\rho_{nn'}$ represents the coherence. 
In our system we consider that the number of photons are much smaller than the number of atoms in the ensemble. Therefore,  $\rho_{00}  \simeq 1$  which  would imply that $\rho_{nn} \simeq \rho_{nn'}\simeq 0$ but $\rho_{n0} \ne 0$. With these approximations  we can write the dynamical equation for the coherence terms as
\begin{align}
\pdv{{\rho_{n0}(z,t)}}{t}+\qty(\mi \delta_{n}+\dfrac{\gamma}{2})\rho_{n0}(z,t)=\dfrac{\mi d_{n0}}{2\hbar}\mathcal E(z,t), \label{Eq:rho}
\end{align}
where the term with $\gamma$ is added phenomenologically to incorporate the environmental effects.

Equation (\eqref{Eq:rho}) characterize the dynamics of the atomic state when it interacts with the external electromagnetic field. In order to solve for the (forward) propagation of light through the  ensemble of atoms we need to consider the effect of the atomic polarization on the electromagnetic field as well which can be written as
\begin{align}
\qty(\pdv{z} +\dfrac{1}{c}\pdv{t})\mathcal{E}(z,t)&=\dfrac{\mi \omega_{_L}}{2\epsilon_{0}c}\mathcal{P}(z,t).\label{Eq:EMWave}
\end{align}
Here $\mathcal P(z,t)$ is the amplitudes of  the atomic polarization which is defined as the expectation value of the aggregate transition electric dipole moment operator $\bm{d}$ of the entire ensemble,
\begin{align}
P(z,t) &=\mathcal{N} \, \text{Tr}[\rho \, \bm{d}]\nonumber\\
&=\mathcal{N} \left(\sum_{n=1}^N d^{*}_{n0}\rho_{n0}e^{-\mi (\omega_{L}t-kz)}+  \sum_{n=1}^N d_{n0}\rho^*_{n0}e^{\mi (\omega_{L}t-kz)}\right),\label{pd}
\end{align}
where $\mathcal{N}$ is the density of atoms. The atomic polarization is driven by the external electric field. Therefore, the time dependence of the polarization and the electric field is the same. Hence, we can write
\begin{align}\label{Eq:Polarization}
P(z,t)&=\dfrac{1}{2}\mathcal{P}(z,t)e^{-\mi (\omega_{L}t- kz)}+\text{cc}.
\end{align}
Comparing Eq.~\eqref{Eq:Polarization} with Eq.~(\ref{pd}) results in
\begin{align}
\mathcal{P}(z,t)=2\mathcal{N} \sum_{n=1}^N d^{*}_{n0}\rho_{n0}.
\end{align}

Equations (\eqref{Eq:rho}) and (\eqref{Eq:EMWave}) together describe the propagation of light through an atomic ensemble~\cite{Meystre2013,Sangouard2007}. We are interested in finding the state of light at time $t$ and position $z$ which requires solving these two equations simultaneously. We can obtain the solution by taking the Fourier transform of $\rho$, $\mathcal{P}$ and $\mathcal{E}$ in the time domain. This would yield
\begin{align}
  \tilde{\rho}_{n0}(z,\omega) &=\dfrac{\mi d_{n0}}{2\hbar}\dfrac{\tilde{\mathcal E}(z,\omega)}{\mi \qty(\omega+\delta_n)+(\gamma/2)},\\ \label{af}
  \mathcal{P}(z,\omega) & = 2 \mathcal{N} \sum_{n} \dfrac{\mi \abs{d_{n0}}^{2}}{2\hbar}\dfrac{\tilde{\mathcal E}(z,\omega)}{\gamma/2+\mi(\delta_{n}+\omega)},
\end{align}
and propagation equation for $\tilde{\mathcal{E}}(z,\omega)$,
\begin{align}
\qty(\pdv{z} +\dfrac{\mi \omega}{c})\tilde{\mathcal{E}}(z,\omega)&=\dfrac{\mi \omega_{L}}{2\epsilon_{0}c}\mathcal{P}(z,\omega).
\end{align}
 
From here it is straight forward to solve for $\tilde{\mathcal{E}}(z,\omega)$ which reads

\begin{align}
\tilde{\mathcal{E}}(z,\omega)=\tilde{\mathcal{E}}(0,\omega)e^{-\mathcal{D}z} \label{leq}
\end{align}
where $\mathcal{D}$ and $g$ are defined as
\begin{align}
\mathcal{D} = \sum_{n} \dfrac{g}{\gamma/2 + \mi(\delta_{n}+\omega)}+\dfrac{\mi \omega}{c}, \quad g=\mathcal{N} \dfrac{\abs{d}^{2}\omega_{L}}{2\hbar\epsilon c}.\label{Eq:D}
\end{align}
Clearly the first term in $\mathcal{D}$ is a comb where the Lorentzian distributions in $\omega$ are placed at $\delta_n$. The second term results in a position dependent phase where $\omega/c = k$. Inverse Fourier transform of $\tilde{\mathcal{E}}(z,\omega)$  gives the output in the time domain. 

\section{\uppercase{Effect of Doppler broadening on the output electric field}}\label{App:Doppler}
In this Appendix we consider the effect of Doppler broadening on the output light. An atomic ensemble at temperature $T$ contains $p_v$ fraction of  atoms moving with velocity $\bm v$ where $p_v$ has the form of thermal distributions, i.e., $p_v = \exp(-mv^2/2k_BT)$. Here $m$ is the mass of the atom and $k_B$ is the Boltzmann constant. An atom with velocity $\bm{v}$ and wave vector $\bm{k}$ experience the modified frequency of the incoming light from $\omega_L$ to  $\omega_L+ \bm{k} \cdot \bm{v}$.  This would modify the detuning $\delta$ to  $\delta + \bm{k} \cdot \bm{v}$ which shows in  $\mathcal{D}$ in Eq.~\eqref{Eq:D} as
\begin{align}
\mathcal{D}_{\bm v} = \sum_{n} \dfrac{g}{\gamma + \mi(\delta_{n} + \bm{k\cdot v} + \omega)}+\dfrac{\mi \omega}{c}.
\end{align}

From this modified $\mathcal{D}_{\bm{v}}$ we can calculate the output electric field
\begin{align}
\tilde{\mathcal{E}}(z,\omega,\bm{v})=\tilde{\mathcal{E}}(0,\omega)e^{-\mathcal{D}_{\bm v}z}. \label{leqv}
\end{align}
Hence the light emitted from an atom moving with velocity $\bm v$ is $\tilde{\mathcal{E}}(z,\omega,\bm{v})$ which is different from $\tilde{\mathcal{E}}(z,\omega,\bm{v} = 0)$ by only an over all phase which can be calculated by replacing $\omega \to \omega' = \bm{k\cdot v} + \omega$. Thus the additional phase the $\tilde{\mathcal{E}}(z,\omega,\bm{v})$ gets is $\exp{\mi \bm{k}\cdot\bm{v} z/c}$. Since, the velocity distribution is incoherent over the atomic ensemble, we need to add the intensity of light coming from each atom which is independent of the velocity. Thus, the intensity of light in the photon echo is unchanged due to the Doppler effect.

\section{\uppercase{Transitions in Cs atom}}\label{App:Transitions}
In this Appendix we calculate the transitions between the states 5$p^{6}$6s (degenerate ground state) and 5$p^{6}$8p (degenerate excited state). The Hamiltonian of the atom in the presence of an external magnetic field can be written as~\cite{bransden}
\begin{align}
H=H_{_0}+H_{\text{hfs}}+H_{_B},
\end{align}
where $H_0$ is the unperturbed spin-less Hamiltonian plus the fine-structure Hamiltonian of the atom and $H' =  H_{\text{hfs}}+H_{_B}$ is the perturbation. The Hamiltonian $H_{\text{hfs}} = A(\bm{I\cdot J})$ is the hyperfine correction and $H_{_B} = -\bm{(\hat{\mu}_{_J}+\hat{\mu}_{_I})\cdot B} = (g_{_{J}}\mu_{_{B}}J_{z}-g_{_{I}}\mu_{_{N}}I_{z})B$ is the interaction of the atom with the magnetic field. Here $A$  is the hyperfine structure constant, $\bm{I}$ and $\bm{J}$ are nuclear spin operator and total (spin and orbital angular momentum) operator, and $\mu_{_B}$ and $\mu_{_N}$ are the Bohr and nuclear magneton. $g_{_{J}}$ and $g_{_{I}}$ are given as
\begin{align}
\begin{aligned}
g_{_{J}}=1+\dfrac{J(J+1)+S(S+1)-L(L+1)}{2J(J+1)},
\end{aligned}
\end{align}
$g_{_{I}} = 0.7369$  for a Cs atom.

\begin{figure}[t!]
  \subfigure[]{
    \includegraphics[width= 6cm]{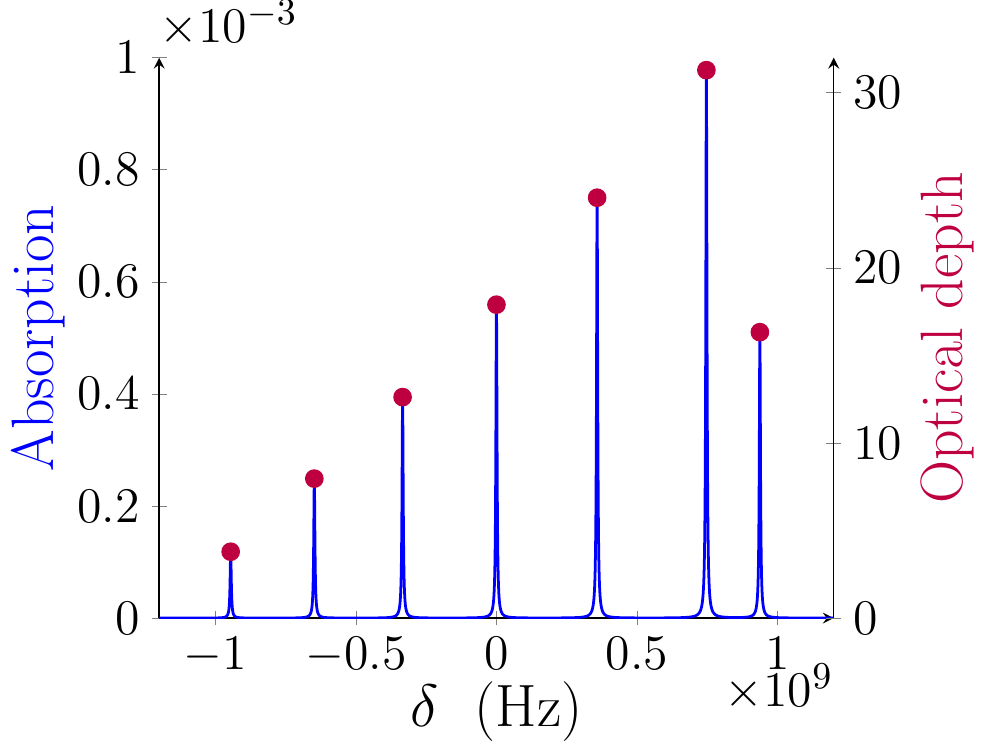}
    \label{fig:Sup+1}} 
\subfigure[]{
  \includegraphics[width= 6cm]{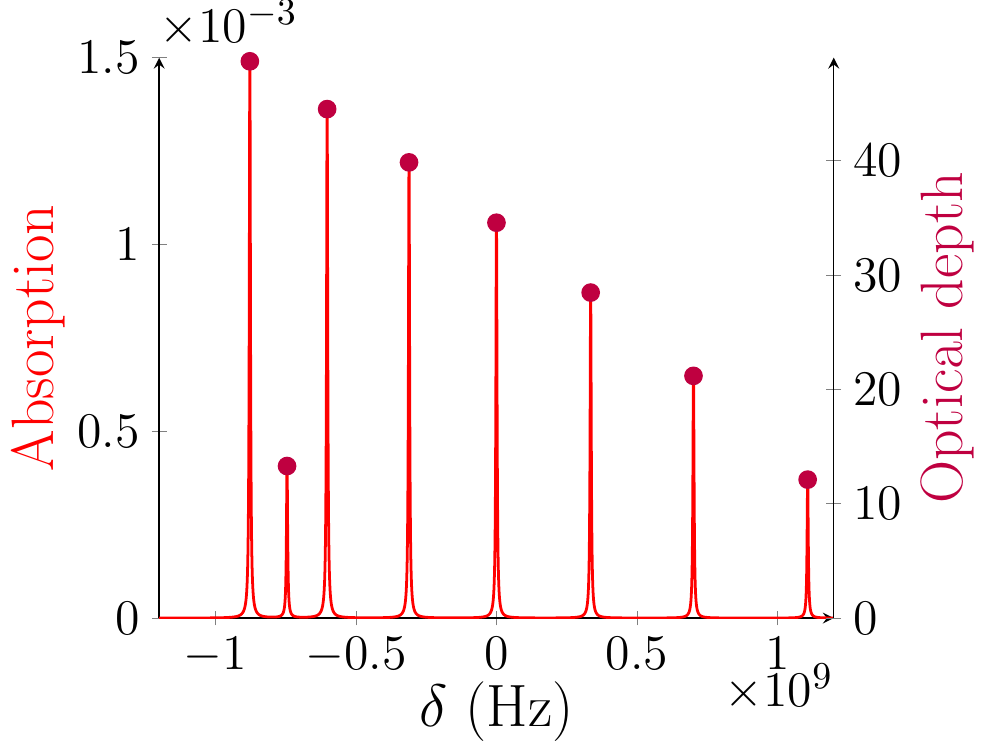}
\label{fig:Sup-1}}\\
\caption[]{(Color online) Numerically calculated absorption coefficients and optical depth for various  transitions for  $\Delta m = +1$ and $\Delta m = -1$ for Cs atom. Here the $x$-axis is the detuning with respect to the incoming light and $y$-axis is the value of the absorption for various transitions. The natural line width of  each transition is taken to be 5MHz. }
 \end{figure}

We are interested in only the states 5$p^{6}$6s and  5$p^{6}$8p which are the eigenstates of the Hamiltonian $H_0$. The energy difference between these states is $773.20996$THz which lies in the optical frequency range~\cite{Liu2000,Safronova1999}. The 5$p^{6}$6s states which we call ground states have $16$-fold degeneracy and the 5$p^{6}$8p energy level (excited states) has 32-fold degeneracy. The perturbation $H'$ helps lift the degeneracy in these states which we exploit for AFC purposes.

To construct the AFC in the Cs atom we need to find the absorption spectrum of the transitions between 5$p^{6}$6s and  5$p^{6}$8p in the presence of the perturbative corrections $H'$. For that purpose we write the Hamiltonian $H = H_0 + H'$ in the basis $\ket{F,M_{F},I,J}$ which is the simultaneous eigenbasis of $\bm{I}$, $\bm{J}$ $\bm F$ and $F_z$. Here $\bm F = \bm I + \bm J$. We numerically solve the Hamiltonian $H$ to find the eigenvectors and the eigenvalues. 

The absorption amplitudes are calculated from  the imaginary part of electric susceptibility $\chi_e$ where the atomic polarization $\mathcal P = \epsilon_0 \chi_e {\mathcal{E}}$~\eqref{af}. Therefore the absorption coefficients reads 
\begin{align}
  \text{Im} \chi_e = \frac{\mathcal{N}}{\hbar \epsilon_0} \sum_n \frac{ |d_{n_en_0}|^2\gamma}{\gamma^2 +(\delta_n + \omega)^2}.
\end{align}
Optical depth $\alpha$ is defined as
\begin{align}
  \alpha = \dfrac{\mathcal{N}\abs{d}^2 \omega_L L}{2\hbar\epsilon_0 c \gamma}.
\end{align}
Here $L$ is the length of the cell which we have taken to be $5cm$, $\gamma = 5\text{MHz}$ is the decay rate of the Cs transition at 100k temperature~~\cite{Liu2000}. $\mathcal N$ is the number density of the Cs atoms in the ensemble which is taken to be $10^{18}m^{-3}$. The magnetic field strength is taken to be $0.1$T. The dipole matrix elements $d_{n_en_0} = \bra{n_e}\bm d\ket{n_0}$, where $\ket{n_e}$ is one of the excited states and $\ket{n_0}$ is one of the ground states which can easily be calculated numerically. In Figs.~\ref{fig:Sup+1} and~\ref{fig:Sup-1} we show the absorption coefficients and the optical depths for the $\Delta m = \pm 1$ transitions.

\section{\uppercase{Optimum efficiency for quantum memory}}\label{App:Efficiency}

The transmission of electric field through an absorptive medium can be written as an exponential decay of the amplitude, i.e., 
\begin{align}
\mathcal{E}_f(L,t)&=e^{-\alpha L/2} \mathcal{E}_f(0,t),
\end{align}
where $\mathcal{E}_f(L,t)$ is the amplitude of the electric field at time $t$ and position $L$, and $\alpha$ is the absoption coefficient of the medium. Here we have assumed $L$ is so small that the time  $L/c$ is neglected. The efficiencies of the backward and forward photon echo in AFC can be determined by the same absorption coefficient $\alpha$. The efficiency of the forward photon-echo $\eta_f$ and the backward photon-echo $\eta_b$ is given by~\cite{Afzelius2009}
\begin{align}
  \eta_f(L) = \epsilon (\alpha L)^2 \exp(-\alpha L),\label{Eq:Eta-f}\\
  \eta_b(L) = \epsilon [1-\exp(-\alpha L)]^2,
\end{align}
where $\epsilon$ is the proportionality constant which is  same for the two efficiencies. 

\begin{figure}[!h]
\includegraphics[width=6cm]{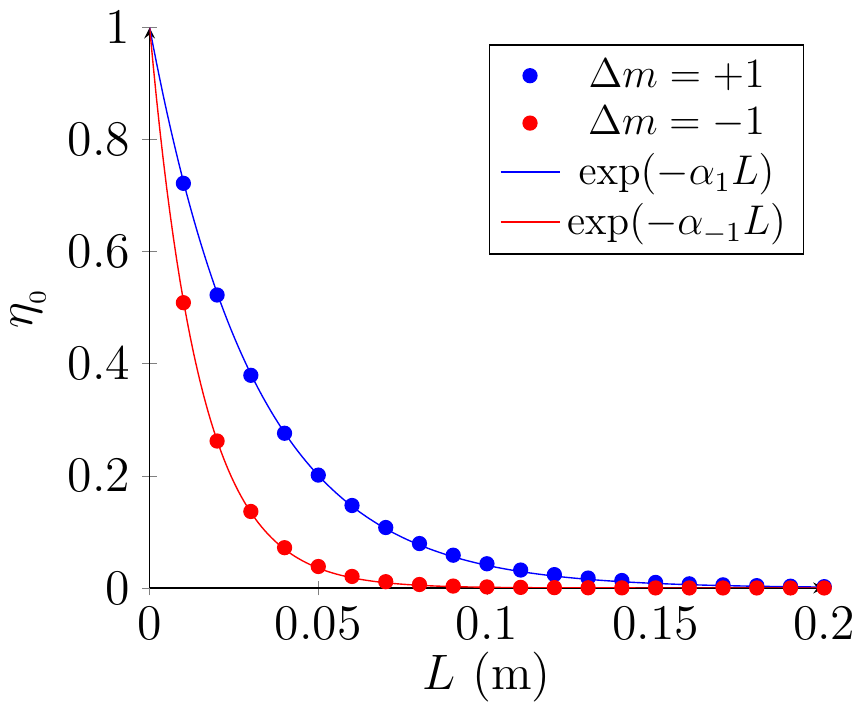}
  \caption{(Color online) Data and the curve fits for, efficiency vs length at 0k. The absorption coefficients obtained from curve fitting are $\alpha_1=32.09$ and $\alpha_{-1}=66.77$}\label{fig:eff0}
\end{figure}

Since in the I-AFC different transitions have different absorption coefficients, we can assume that the total medium have an average coefficient $\alpha_{\Delta m}$ for $\Delta m = \pm1$. In Fig.~\ref{fig:eff0} we plot the numerically calculated values of $\eta_0 = \mathcal{E}_f(L,t)/\mathcal{E}_f(0,t)$ for $\Delta m = \pm 1$ and calculate the coefficient $\alpha_{\pm1}$ from the data. From here we found that $\alpha_1=32.09$ and $\alpha_{-1}=66.77$.

\begin{figure}
\includegraphics[width=6cm]{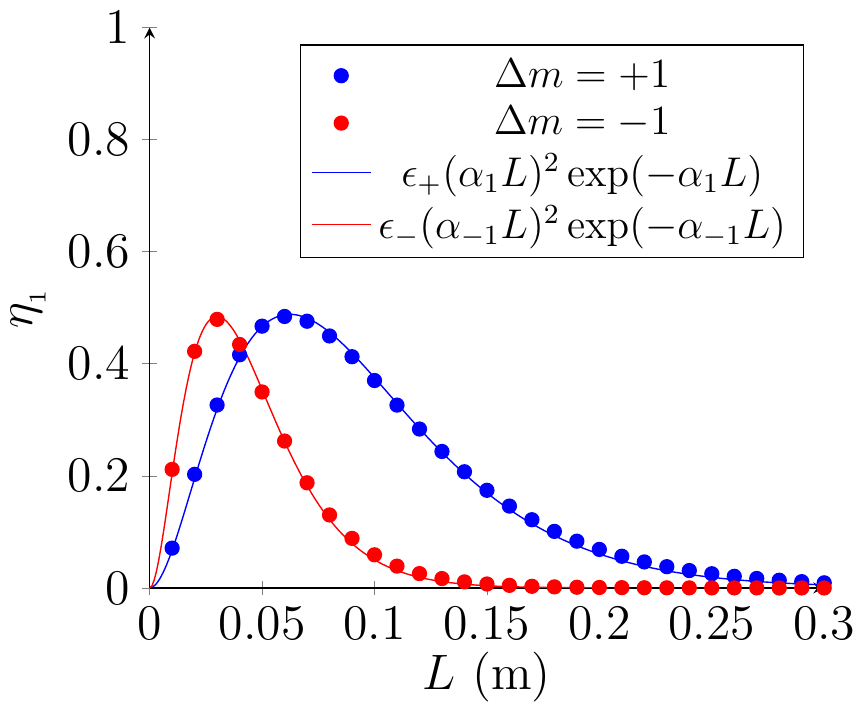}
\caption{(Color online) Data and the curve fits for, efficiency vs length at 0k.}\label{fig:eff1}
\end{figure}
\begin{figure}
\includegraphics[width=6cm]{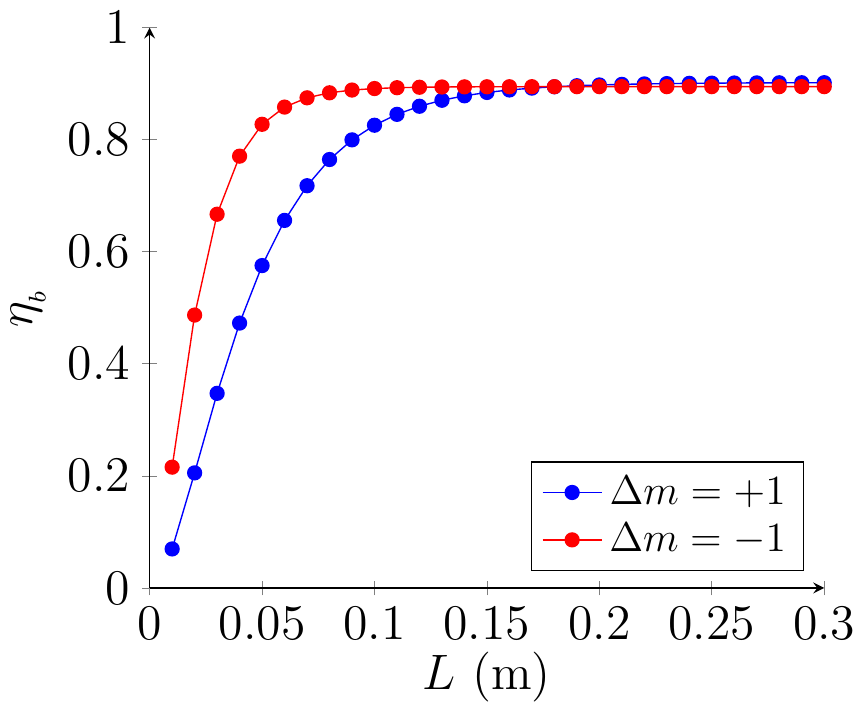}
\caption{(Color online) Backward mode efficiency vs length at 0k.}\label{fig:back}
\end{figure}

In Fig.~\ref{fig:eff1} we plot the numerically calculated efficiency for the forward photon echo $\eta_f$ for I-AFC. We can see that the efficiency $\eta_f$ satisfies the relation Eq.~\eqref{Eq:Eta-f} with unknown $\epsilon$ which can be calculated from this numerical data. From this numerical simulation we found $\epsilon_+ = 0.902$ and $\epsilon_- = 0.895$ for $\Delta m = \pm 1$ transitions. With this numerically calculated value of $\epsilon_\pm$ we plot the efficiency of the backward photon echo as a function of $L$ in Fig.~\ref{fig:back}. We can see that the optimum backward efficiency achieved in the I-AFC in Cs atoms is close to $90\%$ for both the transitions. 


%

\end{document}